\documentstyle[prb,aps,floats]{revtex}

\begin{document}
\twocolumn[\hsize\textwidth\columnwidth\hsize\csname @twocolumnfalse\endcsname

\title{Universality of the Hohenberg-Kohn functional}
\author{Arno Schindlmayr$^{\rm a)}$}
\address{Fritz-Haber-Institut der Max-Planck-Gesellschaft, Faradayweg 4--6,
  14195 Berlin-Dahlem, Germany}
\date{Comment submitted to the American Journal of Physics}
\maketitle

\begin{abstract}
\end{abstract}

]\narrowtext

\noindent
In a recent article\cite{Nea98} H.~L.\ Neal presents a pedagogical approach to
density-functional theory\cite{Hoh64} in the formulation of Kohn and
Sham,\cite{Koh65} which is still largely ignored in undergraduate teaching
despite its enormous significance in many branches of physics, by discussing
the application to one-dimensional two-particle systems. In this context Neal
derives an analytic expression for the Hohenberg-Kohn functional $F[\rho]$,
given by Eq.\ (30) of the original paper, that he suggests is exact for all
systems with the harmonic interaction $u(x_1,x_2) = k (x_1-x_2)^2 / 2$. The
purpose of this comment is to refute this claim for arbitrary external
potentials and to point out that the functional used in Ref.\
\onlinecite{Nea98} really constitutes an approximation in the same spirit as
the local-density approximation.

Exploiting the universality of the Hohenberg-Kohn functional, Neal calculates
\begin{equation}
F = E - \int\!\! v(x) \rho(x) \,dx
\end{equation}
for an analytically solvable model of two coupled harmonic oscillators. He
then rewrites the total energy $E$ and the external potential $v$ on the
right-hand side in terms of the density $\rho$, using exact relations that are
only valid at the ground state of the particular model, however. This
substitution hence replaces the explicit dependence on the external potential
by a system-specific energy surface and fails to produce a universal
functional. In particular, the minimum of the total energy obtained in this
way in general differs from the true ground state. This subtle point is best
seen if the resulting functional $F[\rho]$ is directly inserted into the
variational expression
\begin{equation}
\frac{\delta}{\delta \rho(x)} \left( F[\rho] + \int\!\! \left( v(x) - \mu
  \right) \rho(x) \,dx \right) = 0
\end{equation}
that determines the ground-state density for arbitrary external
potentials.\cite{Hoh64} The Lagrange multiplier $\mu$ enforces the proper
normalization. In the notation of Ref.\ \onlinecite{Nea98}, which also
provides $\delta F[\rho] / \delta \rho(x)$ [Eq.\ (31)], one thus obtains
\begin{equation}
\rho(x) = \rho_0 \exp\left( -\frac{4 \omega (v(x) - v(0))}{\hbar \omega_0
    (\omega_0 + \omega)} \right) \label{eq:density}
\end{equation}
and hence recovers the specific relation between $v$ and $\rho$ that was
previously employed to construct the functional $F[\rho]$ in the first place.
By design, this expression is exact for the coupled harmonic oscillators. For
all other external potentials, however, Eq.\ (\ref{eq:density}) constitutes an
approximation to the ground-state density that does not coincide with the
minimum of the true total-energy surface.

The Kohn-Sham formalism requires $F[\rho]$ as an auxiliary quantity to define
\begin{equation}
F_r[\rho] = F[\rho] - T_r[\rho] ,
\end{equation}
where $T_r[\rho]$ denotes the kinetic energy of the noninteracting reference
system. In the usual terminology, $F_r[\rho]$ contains the Hartree and the
exchange-correlation part of the total energy, its derivative $\delta
F_r[\rho] / \delta \rho(x)$ contributes to the effective potential that
appears in the single-particle Kohn-Sham equations. It follows from the above
argument that this term is not treated exactly in Ref.\ \onlinecite{Nea98},
because the Hohenberg-Kohn functional as well as $T_r[\rho]$ are constructed
for a particular system and hence are not free from an implicit dependence on
the harmonic-oscillator potential. The Hartree and exchange-correlation
potential is thus modelled by that of two coupled harmonic oscillators. In
this sense the approach is analogous in spirit to the local-density
approximation\cite{Koh65} widely used in atomic and solid-state physics, which
similarly replaces the exact exchange-correlation energy by that of a
homogeneous electron gas with the same local density.

\begin{table}[b]
\caption{Comparison of the exact total energy $E$, the approximate Kohn-Sham
  solution $E_{\rm KS}$ and a variational estimate $E_{\rm var}$ for two
  different confining potentials.\label{tab:energy}}
\begin{tabular}{cccc}
case & $E$ & $E_{\rm KS}$ & $E_{\rm var}$\tablenote{From Ref.\
  \onlinecite{Nea98}.} \\
\hline
2 & $1.962\,81$ & $1.963\,54$ & $1.967\,66$ \\
3 & $2.981\,49$ & $2.981\,56$ & $2.981\,68$ \\
\end{tabular}
\end{table}

Much as the local-density approximation is successful for weakly inhomogeneous
systems, so Neal's effective potential may be applied to one-dimensional
two-particle systems with the harmonic interaction in general confining
potentials. Indeed, Ref.\ \onlinecite{Nea98} presents meaningful results for
several potential wells with different shapes. For a quantitative analysis we
carefully reexamine the examples discussed in the original paper. We
diagonalize the Hamiltonians using a basis of noninteracting
harmonic-oscillator eigenfunctions. In this way all matrix elements can be
calculated either analytically or by a numerically exact Gauss-Hermite
quadrature. The results for case 2: $v(x) = \alpha |x|$ and case 3:
$v(x) = \alpha \exp(\beta x^2)$, converged with respect to the number of basis
functions, are more accurate and hence differ from those quoted in Ref.\
\onlinecite{Nea98}. We have set $\alpha = 1.0$, $\beta = 0.1$ and $k = 1.0$.
In Table \ref{tab:energy} we contrast the exact total energy $E$ with the
Kohn-Sham solution $E_{\rm KS}$. Although small, the deviation is genuine. For
comparison we also quote the estimate $E_{\rm var}$ from a two-parameter
variational wave function given in Ref.\ \onlinecite{Nea98}. The two
approximate schemes yield similar small errors if the quantum well resembles
a harmonic potential near the origin (case 3), otherwise the variational wave
function is less appropriate and the effective potential gives better
agreement with the exact solution (case 2).

In summary, although not exact, Neal's scheme yields good approximate results
for confined two-particle systems with the harmonic interaction in one
dimension. Furthermore, the procedure closely follows the spirit of the widely
used local-density approximation and is hence of additional pedagogical value.
We hope that it will find recognition and contribute towards making
density-functional theory more accessible to students.

\end{document}